\documentclass[12pt]{article}
\def\tilde{\widetilde}
\def\hat{\widehat}
\newcommand\text[1]{\hbox{\rm #1}}
\input tcilatex
\begin{document}

\title{\rightline{\small TAUP-2610-2000}  \rightline{\small hep-ph/0001182} 
Production of fermions \\
in models of string cosmology}
\author{Ram Brustein ${}^1$, Merav Hadad ${}^2$ \\
{%
\hbox{ \small $1$ Department of Physics, Ben-Gurion University, 
Beer-Sheva 84105, Israel}} \\
{%
\hbox{\small $2$ School of Physics and Astronomy, 
Tel Aviv University, Tel Aviv 69978, Israel }} \\
{\hbox{\small email: ramyb@bgumail.bgu.ac.il, meravv@post.tau.ac.il} }}
\date{}
\maketitle

\begin{abstract}
Production of spin 1/2 fermions and gravitinos by the standard mechanism of
amplification of quantum fluctuations during the dilaton-driven inflation
phase in models of string cosmology is highly suppressed. Constraints on
string cosmology models from gravitational production of gravitinos,
contrary to expectations, are similar to constraints on other models.
\end{abstract}

\setlength\baselineskip{18pt}

\section{Introduction}

It is well known that particles are produced in cosmological backgrounds by
amplification of vacuum fluctuations during inflationary epochs \cite{mukh}.
In models of string cosmology \cite{pbb,review}, most of the produced
particle spectra were found to be very different from the spectra in other
cosmological models. First, in string cosmology models, gravitational waves
may be generated in a substantial amount, in contrast to slow-roll inflation
which predicts a ``desert''\cite{bggv,maggiore}. Furthermore, axions and
other scalar particles have phenomenologically interesting spectra [6-10].
Finally, it is important to note that in addition to scalar particles and
gravitational waves, photons can be produced \cite{photon,Lemoine:1995}. In
general relativity, gauge field perturbations are not amplified in
conformally flat cosmological backgrounds (even if inflationary) because of
the scale invariant coupling of gauge fields in four dimensions. However, in
string cosmology, the presence of a time-dependent dilaton prefactor in
front of the gauge-field kinetic term breaks scale invariance and results in
photon production which may even lead to the creation of macroscopic
magnetic fields \cite{photon,Lemoine:1995}.

We study fermion production by the standard amplification of quantum
fluctuation in models of string cosmology. Although expectations were that
fermions are copiously produced, we have found that production of
spin 1/2 and spin 3/2 fermions in models of string cosmology during the
dilaton-driven inflationary (DDI) phase is highly suppressed.

It is known that massless spin 1/2 fermions are not amplified in conformally
flat cosmological backgrounds \cite{parker}. This is a consequence of their
conformally invariant interactions. We show that in models of string
cosmology the presence of a time-dependent dilaton prefactor in front of the
fermion kinetic term does not result in amplification of fermionic
perturbations. The spin-3/2 gravitino, the supersymmetric partner of the
graviton, can be produced in conformally flat cosmological backgrounds,
since gravitinos do not have conformally invariant couplings. It was
recently shown that if gravitinos have vanishing zero-temperature, flat
background mass, then they are not produced [14-18]. We show that, in models
of string cosmology, the appearance of the dilaton prefactor in front of the
gravitino kinetic terms does not result in amplification of massless
gravitino perturbations, and therefore that gravitino production during DDI
is negligible.

We consider fermion production in models of string cosmology which realize
the pre-big-bang scenario \cite{pbb}. In this scenario the evolution of the
universe starts from a state of very small curvature and coupling and then
undergoes a long phase of DDI and at some later time smoothly joins standard
radiation dominated (RD) cosmological evolution, thus giving rise to a
singularity free inflationary cosmology. Particles are produced during the
period of DDI by the standard mechanism of amplification of quantum
fluctuations. We assume throughout an isotropic and homogeneous four
dimensional flat universe, described by a FRW metric. As in \cite{bh}, the
mass of the fermions is assumed to vanish in the early universe, i.e. in the
dilaton driven period, and to take a nonzero value from the start of the RD
era and on. All fields, except the dilaton and metric, are assumed to have a
trivial vacuum expectation value during the inflationary phase.

Fermion production, and in particular gravitino production, has the
potential of placing severe constraints on models of string cosmology. If it
turned out that gravitinos are maximally produced during DDI as
gravitons are (a reasonable assumption) then after supersymmetry is broken,
gravitinos can become by far the dominant energy component in the universe.
The universe would become matter dominated very early, and the resulting
cosmology will be very different than the standard RD cosmology.
The issue of gravitino production during the evolution 
following the DDI phase is still unresolved, and we do not discuss thermal 
production which may turn out to put tight constraints on models of 
string cosmology, as it does for other models of inflation.

\section{Spin 1/2 fermions}

We show here that massless spin 1/2 particles are not produced during DDI in
models of string cosmology. We first explicitly show using their equations
of motion that they are not produced, and then show that they are not
produced even for a space and time dependent dilaton 
by showing that the low energy string effective action of massless
spin 1/2 fermions can be transformed into a conformally invariant action.

In the low energy effective action, a dilaton prefactor appears in front of
the kinetic term, 
\begin{equation}
\mathcal{A}=\int d^4x\sqrt{-g}e^{l\phi }\left\{ \frac i2\left[ \bar \psi
\hat\gamma ^\mu D_\mu \psi -\left( D_\mu \bar \psi \right) \hat\gamma ^\mu \psi
\right] -M\bar \psi \psi \right\} .  \label{af}
\end{equation}
Greek letters denote space-time indices and Latin letters denote
tangent-space indices. Gamma matrices and the group generators for spin 1/2
irreducible representations with curved indices are defined by 
\begin{equation}
\hat \gamma ^\mu =e_a^\mu (x)\gamma ^a\hbox{\rm  , }\hat \Sigma ^{\lambda
\nu }=\Sigma ^{ab}e_a^\lambda e_b^\nu
\end{equation}
where $e_\alpha ^\mu $ is the vierbein. $D_\mu $ is the covariant derivative
with respect to the spinorial structure 
\begin{equation}
D_\mu =\partial _\mu +\Gamma _\mu =\partial _\mu +\frac 12g_{\lambda \delta
}\Gamma _{\nu \mu }^\delta \hat \Sigma ^{\lambda \nu }.
\end{equation}
Fermions have $l=-1$, but we keep parameter $l$ general, since our results
do not depend on its specific value.

Variation of the action with respect to $\bar \psi $ yields the Dirac
equation, 
\begin{equation}
i\hat \gamma ^\mu D_\mu \psi -M\psi +\frac i2l\hat \gamma ^\mu \partial _\mu
\phi \psi =0.  \label{emf}
\end{equation}

In the simplified model of background evolution we adopt, the evolution of
the universe is described by (conformal) time dependent dilaton $\phi (\eta
)$ and a spatially flat FRW metric, in which the line element can be written
as $d^2s=a^2(\eta )\left( d^2\eta -d^2\vec x\right) .$ Here $\vec
x=(x^1,x^2,x^3)$ are comoving space coordinates and $x^0\equiv \eta $ is the
conformal time. Thus, the vierbein can be written as $e_\mu ^a=a\delta _\mu
^a,$ $e_a^\mu =a^{-1}\delta _a^\mu $. In this case, the equation of motion (%
\ref{emf}) reduces to 
\begin{equation}
\gamma ^a\partial _a\psi +\frac 12\frac{a^{\prime }}a\vec \gamma \vec \Sigma
\psi +iMa\psi ++\frac l2\partial _0\phi \gamma ^0\psi =0,
\end{equation}
where $\vec \Sigma =(\Sigma ^{01},\Sigma ^{02},\Sigma ^{03}).$ Choosing a
gamma-matrix representation such that 
\begin{equation}
\gamma ^0=\left( 
\begin{array}{cc}
I & 0 \\ 
0 & -I
\end{array}
\right) ;\vec \gamma =\left( 
\begin{array}{cc}
0 & \vec \sigma \\ 
-\vec \sigma & 0
\end{array}
\right) ,
\end{equation}
the field $\psi$ satisfies the following equations of motion 
\begin{eqnarray}
\left( \partial _0+iMa+\frac l2\partial _0\phi +\frac 32\frac{a^{\prime }}%
a\right) \psi _1+\sigma ^i\partial _i\psi _2 &=&0  \label{emf1} \\
\left( \partial _0-iMa+\frac l2\partial _0\phi +\frac 32\frac{a^{\prime }}%
a\right) \psi _2+\sigma ^i\partial _i\psi _1 &=&0.  \nonumber
\end{eqnarray}
Introducing the field $\chi $, 
\begin{equation}
\psi (\eta ,\vec x)=a^{-3/2}e^{-\frac l2\phi }\chi (\eta ,\vec x),
\end{equation}
equations (\ref{emf1}) reduce to 
\begin{eqnarray}
\left( \partial _0+iMa\right) \chi _1+\sigma ^i\partial _i\chi _2 &=&0
\label{emf2a} \\
\left( \partial _0-iMa\right) \chi _2+\sigma ^i\partial _i\chi _1 &=&0.
\label{emf2b}
\end{eqnarray}
Substituting (\ref{emf2a}) into (\ref{emf2b}) and vice-versa, and writing $%
\chi $ as $\chi (\eta ,\vec x)=\chi (\eta )e^{ikx}$, we arrive at 
\begin{equation}
\left( \partial _0^2+k^2+M^2a^2\pm iMa\right) \left( 
\begin{array}{c}
\chi _1 \\ 
\chi _2
\end{array}
\right) =0.
\end{equation}
Finally, the equation of massless spin 1/2 fermions is reduced to the
equation in flat space-time and therefore these particles are not produced.

Let us turn now to the massless low energy effective action of spin 1/2
particles and show that it is conformally invariant, even for 
a space and time dependent dilaton and scale factor. Under the conformal
transformation 
\begin{equation}
g^{\mu \nu }\rightarrow \tilde g^{\mu \nu }=\Omega ^2g^{\mu \nu },
\end{equation}
vierbeins, gamma matrices and covariant derivatives transforms as follows, 
\[
e_a^\mu \rightarrow \tilde e_a^\mu =\hbox{\rm  }\Omega e_a^\mu 
\]
\[
\hat \gamma ^\mu \rightarrow \tilde{\hat\gamma}{}^\mu =\Omega \hat \gamma ^\mu 
\]
\[
D_\mu \rightarrow \tilde D_\mu =D_\mu +\frac 12\Sigma ^{ab}e_a^\nu g_{\rho
\nu }e_b^\rho \frac \partial {\partial x^\mu }\left( \ln \Omega \right) ,
\]
and $\psi $ transform as follows 
\[
e^{l\phi (x)}\rightarrow e^{l\tilde \phi (x)}=\varphi \left( x\right)
e^{l\phi (x)}
\]
\[
\psi \rightarrow \tilde \psi =\Omega ^{3/2}(x) 
\varphi ^{-1/2}\left( x\right)\psi.
\]
The low energy effective action of massless spin 1/2 particles (\ref{af})
transforms into 
\begin{equation}
\mathcal{A}\rightarrow \tilde\mathcal{A}=\int d^4x\sqrt{-\tilde g}%
e^{l\tilde \phi }\frac 12i\left[ \bar{\tilde {\psi}}
 \tilde{\hat\gamma}{}^\mu \tilde D_\mu
\tilde \psi -\left( \tilde D_\mu \bar{\tilde{\psi}} \right) 
\tilde{\hat\gamma}{}^\mu \tilde \psi \right] =
\end{equation}
\[
\int d^4x\sqrt{-g}e^{l\phi }\left[ \frac 12i
\left( \bar \psi \hat \gamma^\mu D _\mu \psi 
-(D _\mu \bar \psi )\hat \gamma ^\mu \psi \right)
\right] +
\]
\[
\int d^4x\sqrt{-g}e^{l\phi }\left[ \frac 14i\bar \psi \left( \hat \gamma
^\mu \Sigma ^{ab}-\Sigma ^{ab}\hat \gamma ^\mu \right) \psi e_a^\nu g_{\rho
\nu }e_b^\rho \frac \partial {\partial x^\mu }\left( \ln \Omega \right)
\right] .
\]
Since $\hat \gamma ^\mu \Sigma ^{ab}-\Sigma ^{ab}\hat \gamma ^\mu $ is
antisymmetric in $(a,b)$ and $e_a^\nu g_{\rho \nu }e_b^\rho $ is symmetric
in $(a,b)$, it follows that
\[
\left( \hat\gamma ^\mu \Sigma ^{ab}-\Sigma ^{ab}\hat\gamma ^\mu \right) e_a^\nu
g_{\rho \nu }e_b^\rho =0.
\]
Thus, the low energy effective action (\ref{af}) is conformally invariant in
the massless limit. In other words, fluctuations of massless spin 1/2
fermions are not amplified in a conformally flat background, even if
inflationary, in agreement with our previous calculation.

Action (\ref{af}) is valid during the DDI period, and since the fermions are
assumed to be massless during that period they are not produced. Following
that period, during the string phase, fermions are assumed massless, and it
seems reasonable to assume that the fermions are not produced during the
string phase as well (but see below), if so they may be produced only during
the RD and MD epochs, by thermal and non-thermal production. Since in the RD
and MD era the dilaton is assumed to be fixed and constant, we conclude that
the spectra of the produced spin 1/2 particles in string cosmology models
are similar to those in other models, and therefore the cosmological
constraints and observable features are common.

\section{Gravitinos - spin 3/2 fermions}

In this section we show that the appearance of a time-dependent dilaton
prefactor in the low energy effective action does not change the
amplification of fluctuation of massless gravitinos. Before studying the
properties of the action, it is useful to recall a few known facts about
gravitinos in conformally flat cosmological backgrounds.

Massive gravitinos in conformally flat space time are described by four
Majorana spinors $\psi ^\mu $, satisfying the equations \cite{Kallosh:1999,rioto} 
\begin{eqnarray}
\ &&\left[ i\gamma ^0\partial _0+i\frac{5\dot a}{2a}\gamma ^0-ma+k\gamma
^3\right] \psi _{3/2}=0,  \nonumber \\
\ &&\left[ i\gamma ^0\partial _0+i\frac{5\dot a}{2a}\gamma ^0-ma+k\left(
A+iB\gamma ^0\right) \gamma ^3\right] \psi _{1/2}=0,  \label{emg} \\
\ A &=&\frac 1{3\left( \frac{\dot a^2}{a^4}+m^2\right) ^2}\left[ 2\frac{%
\ddot a}{a^3}\left( m^2-\frac{\dot a^2}{a^4}\right) +\frac{\dot a^4}{a^8}%
-4m^2\frac{\dot a^2}{a^4}+3m^4-4\frac{\dot a}{a^3}\dot mm\right]  \nonumber
\\
\ &&B=\frac{2m}{3\left( \frac{\dot a^2}{a^4}+m^2\right) ^2}\left[ 2\frac{%
\ddot a\dot a}{a^5}-\frac{\dot a^3}{a^6}+3m^2\frac{\dot a}{a^2}+\frac{\dot m%
}{ma}\left( m^2-\frac{\dot a^2}{a^4}\right) \right] .  \nonumber
\end{eqnarray}
Here $\psi _{1/2}$ and $\psi _{3/2}$ are defined such that 
\begin{eqnarray}
\psi ^0 &=&\sqrt{\frac 23}~c~{\hat \gamma }_3~d~{\hat \gamma }_1~\psi _{1/2}
\\
\psi ^1 &=&\frac 1{\sqrt{6}}~\psi _{1/2}+\frac 1{\sqrt{2}}~\psi _{3/2} \\
\psi ^2 &=&{\hat \gamma }^2{\hat \gamma }_1\left( \frac 1{\sqrt{6}}~\psi
_{1/2}-\frac 1{\sqrt{2}}~\psi _{3/2}\right) \\
\psi ^3 &=&\sqrt{\frac 23}\left( d-{\hat \gamma }^3\right) {\hat \gamma }%
_1~\psi _{1/2}
\end{eqnarray}
and 
\begin{eqnarray}
c &=&\frac 1{3a\left( \frac{\dot a^2}{a^4}+m^2\right) }\left[ \left( -2\frac{%
\ddot a}{a^3}+\frac{\dot a^2}{a^4}-3m^2\right) \gamma ^0+2i\frac{\dot m}%
a\right] \\
d &=&\frac{|\vec k|}{a^2\left( \frac{\dot a^2}{a^4}+m^2\right) }\left( i%
\frac{\dot a}{a^2}\gamma ^0+m\right),
\end{eqnarray}
and $\vec k=\left( 0,0,k\right)$.
The fields $\psi _{1/2}$ and $\psi _{3/2}$ correspond to the $\pm 1/2$ and 
$\pm 3/2$ helicity states in the flat limit.

In the massless limit, the field equation of motion in (\ref{emg}) becomes 
\begin{eqnarray}
\ &&\left[ i\gamma ^0\partial _0+i\frac{5\dot a}{2a}\gamma ^0+k\gamma
^3\right] \psi _{3/2}=0  \nonumber \\
\ &&\left[ i\gamma ^0\partial _0+i\frac{5\dot a}{2a}\gamma ^0+k\left(
A+iB\gamma ^0\right) \gamma ^3\right] \psi _{1/2}=0,  \label{emg1} \\
\ &&A=\frac 13\left[ -2\frac{\ddot a}{\dot a}\frac a{\dot a}+1\right] 
\nonumber \\
\ &&B=0.  \nonumber
\end{eqnarray}

The pre-big bang model suggests that the evolution of the scale factor $%
a(\eta )$ is a power dependent function, i.e., 
\begin{equation}
a=a_s\left( \frac \eta {\eta _s}\right) ^\alpha .
\end{equation}
In this case, $A$ becomes a constant; i.e., $A=\left( 2-\alpha \right)
/3\alpha $, and eqs. (\ref{emg1}) become: 
\begin{eqnarray}
\ &&\left[ i\gamma ^0\partial _0+i\frac{5\dot a}{2a}\gamma ^0+k\gamma
^3\right] \psi _{3/2}=0  \label{emg2} \\
\ &&\left[ i\gamma ^0\partial _0+i\frac{5\dot a}{2a}\gamma ^0+\bar k\gamma
^3\right] \psi _{1/2}=0,  \nonumber
\end{eqnarray}
where $\bar k=\left( \frac{2-\alpha }{3\alpha }\right) k.$ As for spin 1/2
particles, one can reduce eqs. (\ref{emg2}) to the field eqs. of motion in
flat space-time with a suitable transformation. Thus, massless gravitino
fluctuation are not amplified in a ``power low'' expanding universe and no
production of gravitinos occurs during the DDI phase.

Next, we show that the low limit effective action of a massless gravitino
can be transformed to the action of massless gravitino in standard
supergravity, even for a space and  time dependent dilaton background.

The low energy effective action of the gravitino in string theory is the
following, 
\begin{equation}
\mathcal{A}=\int d^4xe^{l\phi }\epsilon ^{\mu \nu \rho \sigma }\bar \psi
_\mu \gamma _5 \hat{\gamma}_\nu D_\rho \psi _\sigma +h.c.,  \label{grvac}
\end{equation}
where 
\begin{equation}
~~~~D_\mu =\partial _\mu +\frac 12\omega _{\mu ab}\sigma ^{ab},
\end{equation}
\begin{eqnarray}
\omega _{\mu ab} &=&\frac 12\left( -C_{\mu ab}+C_{ab\mu }+C_{b\mu a}\right) ,
\\
{C^a}_{\mu \nu } &=&\partial _\mu e_\nu ^a-\partial _\nu e_\mu ^a-\frac
1{2M_P^2}{\bar \psi }_\mu \gamma ^a\psi _\nu \sim \partial _\mu e_\nu
^a-\partial _\nu e_\mu ^a.
\end{eqnarray}
Using the transformation $\psi _\sigma =e^{-l\phi /2}\tilde \psi _\sigma ,$
action (\ref{grvac}) becomes 
\begin{equation}
\int d^4x\left[ -\frac 12l\epsilon ^{\mu \nu \rho \sigma }\tilde \psi _\mu
\gamma _5 \hat{\gamma }_\nu \tilde \psi _\sigma \partial _\rho \phi +\epsilon
^{\mu \nu \rho \sigma }\tilde \psi _\mu \gamma _5 \hat{\gamma }_\nu D_\rho \tilde
\psi _\sigma \right] +h.c.\ .  \label{grvac1}
\end{equation}
The first term in (\ref{grvac1}) is canceled by its hermitian conjugate
since using $\gamma ^0\left[ \gamma _5 \hat{\gamma }_\nu \right] ^{+}\gamma
^0=\gamma _5 \hat{\gamma }_\nu $, 
\begin{equation}
\left[ \epsilon ^{\mu \nu \rho \sigma }\bar \psi _\mu \gamma _5 \hat{\gamma }_\nu
\psi _\sigma \partial _\rho \phi \right] ^{\dagger }=
-\epsilon ^{\mu \nu \rho \sigma }
\bar \psi _\mu \gamma _5 \hat{\gamma }_\nu \psi _\sigma \partial_\rho \phi .
\end{equation}
Thus, 
\begin{equation}
\int d^4xe^{l\phi }\epsilon ^{\mu \nu \rho \sigma }\bar \psi _\mu \gamma _5
\hat{\gamma }_\nu D_\rho \psi _\sigma +h.c.\rightarrow \int d^4x\epsilon ^{\mu
\nu \rho \sigma }\bar \psi _\mu \gamma _5 \hat{\gamma }_\nu D_\rho \psi _\sigma
+h.c.
\end{equation}
when $\psi _\sigma \rightarrow e^{-l\phi /2}\psi _\sigma $, showing that
gravitinos are not produced during the DDI phase, since, as we recall from
eq.(\ref{emg2}), there is no production of massless gravitino, and
therefore, they will not be produced in string cosmology, as well.

The mass of the gravitino is assumed to vanish in the early universe, i.e.
in the DDI period, and then to take a nonzero value after the start of the RD
era when supersymmetry is broken. Thus, using the same argument that we 
have used for spin 1/2 particles, 
we conclude that gravitinos can get produced only during the
subsequent RD and MD era.

\section{Summary and Conclusions}

We have studied fermion production during  DDI in string cosmology
models, and found that massless spin 1/2 fermions and gravitinos are not
produced. This unexpected result can be contrasted with boson production. In
particular, naive expectations would be that gravitinos are copiously 
produced. In general, bosons have different spectra
in string cosmology and in standard slow-roll models, but this is not the
case for fermions. Spin 1/2 particles preserves their conformal invariance,
and are not produced during the DDI. They can be produced from the start of
RD era and on, by thermal and non-thermal production, as in other
cosmological models. Similarly, although the action of spin 3/2 particles is
not conformally invariant, it is invariant, in the massless limit, under
''rescaling''. Thus, assuming massless gravitinos in the early universe,
spin 3/2 particles have the same spectrum in string and standard cosmology.

Our results have a limitation; we have neglected production of fermions
during the string phase. As mentioned before, the string phase is a
high-curvature phase of otherwise unknown properties. Usually, one could
overcome lack of knowledge of string phase evolution by using the fact that
fluctuations of fields are frozen when their wavelength is much larger than
the curvature scale during the string phase. However, fermions seem to never
freeze. Thus the unknown properties of the string phase may be important for
fermions production. This is a puzzling issue which we do not understand.
Causality requires that amplitudes of waves whose wavelength is larger than
the cosmological horizon be constant, but fermions seem to evade this
requirement, since they do not seem to feel the cosmological horizon.

Finally, we note that our results may hold for more general string models.
In this paper, as well as previous ones, we have assumed that the dilaton
depends only on time. However the conformal invariance of the massless spin
1/2 action in the low energy effective string action is preserved also for a
space and time dependent dilaton. As a matter of fact it is easy to
eliminate a space and time dependent dilaton from the actions of spin 1/2
and spin 3/2 fermions. Thus our results are valid even in string cosmology
models that involve a space and time dependent dilaton background.

\section*{Acknoledgments}

We thank L. Kofman, M. Lemoine and A. Riotto for discussions, and
A. Linde for useful comments on the manuscript.


\begin{thebibliography}{99}
\bibitem{mukh}  N.D. Birrell and P.C.W. Davies, Quantum fields in curved
space, Cambridge University Press 1984 ;\\V.~F.~Mukhanov, H.~A.~Feldman and
R.~H.~Brandenberger, 
Phys.\ Rept.\ \textbf{215} (1992) 203. 

\bibitem{pbb}  G.~Veneziano, 
Phys.\ Lett.\ \textbf{B265} (1991) 287;\\M.~Gasperini and G.~Veneziano, 
Astropart.\ Phys.\ \textbf{1} (1993) 317 [hep-th/9211021]. 

\bibitem{review}  J.~E.~Lidsey, D.~Wands and E.~J.~Copeland, ``Superstring
cosmology,'' hep-th/9909061. 

\bibitem{bggv}  R.~Brustein, M.~Gasperini, M.~Giovannini and G.~Veneziano, 
Phys.\ Lett.\ \textbf{B361} (1995) 45 [hep-th/9507017]. 

\bibitem{maggiore}  M.~Maggiore, ``Gravitational wave experiments and early
universe cosmology,'' gr-qc/9909001. 


\bibitem{Copeland:1997}  E.~J.~Copeland, R.~Easther and D.~Wands, 
Phys.\ Rev.\ \textbf{D56} (1997) 874 [hep-th/9701082]. 

\bibitem{bh}  R.~Brustein and M.~Hadad, 
Phys.\ Rev.\ \textbf{D57} (1998) 725 [hep-th/9709022]. 

\bibitem{Buonanno:1998}  A.~Buonanno, K.~A.~Meissner, C.~Ungarelli and
G.~Veneziano, 
JHEP \textbf{9801} (1998) 004 [hep-th/9710188]. 

\bibitem{chdm}  R.~Brustein and M.~Hadad, 
Phys.\ Rev.\ Lett.\ \textbf{82} (1999) 3016 [hep-ph/9810526]. 

\bibitem{cmb}  R.~Durrer, M.~Gasperini, M.~Sakellariadou and G.~Veneziano, 
Phys.\ Lett.\ \textbf{B436} (1998) 66 [astro-ph/9806015];\\A.~Melchiorri,
F.~Vernizzi, R.~Durrer and G.~Veneziano, 
Phys.\ Rev.\ Lett.\ \textbf{83} (1999) 4464 [astro-ph/9905327]. 

\bibitem{photon}  M.~Gasperini, M.~Giovannini and G.~Veneziano, 
Phys.\ Rev.\ Lett.\ \textbf{75} (1995) 3796 [hep-th/9504083].


\bibitem{Lemoine:1995}  D.~Lemoine and M.~Lemoine, 
Phys.\ Rev.\ \textbf{D52} (1995) 1955. 

\bibitem{parker}  L.~Parker, 
Phys.\ Rev.\ \textbf{183} (1969) 1057. 

\bibitem{Maroto:1999}  A.~L.~Maroto and A.~Mazumdar, ``Production of spin
3/2 particles from vacuum fluctuations,'' hep-ph/9904206. 

\bibitem{Kallosh:1999}  R.~Kallosh, L.~Kofman, A.~Linde and A.~Van Proeyen,
``Gravitino production after inflation,'' hep-th/9907124. 


\bibitem{rioto}  G.~F.~Giudice, I.~Tkachev and A.~Riotto, 
JHEP \textbf{9908} (1999) 009 [hep-ph/9907510]; \\G.~F.~Giudice, A.~Riotto
and I.~Tkachev, 
JHEP \textbf{9911} (1999) 036 [hep-ph/9911302]. 




\bibitem{Lemoine:1999}  M.~Lemoine, 
Phys.\ Rev.\ \textbf{D60} (1999) 103522 [hep-ph/9908333]. 

\bibitem{Lyth:1999}  D.~H.~Lyth, ``Late-time creation of gravitinos from the
vacuum,'' hep-ph/9912313. 
\end{thebibliography}
\end{document}